\documentclass[12pt]{iopart}

\usepackage{dcolumn}
\usepackage{bm}
\usepackage{subeqn}
\usepackage{iopams}
\usepackage{setstack}
\usepackage{amsfonts}
\usepackage{amssymb}
\usepackage{graphicx}

\begin{document}


\title[Self-consistent dynamic evolution of magnetic islands in the presence of ECCD]
{Self-consistent modeling of the dynamic evolution of magnetic island growth in the presence of stabilizing ECCD}

\author{Ioanna Chatziantonaki$^1$, Christos Tsironis$^{1,2}$, Heinz Isliker$^1$ and Loukas Vlahos$^1$}

\address{$^1$ Department of Physics, Aristotle University of Thessaloniki, 54 124 Thessaloniki, Greece}

\address{$^2$ School of Electrical and Computer Engineering, National Technical University of Athens,
157 73 Athens, Greece}

\ead{ctsironis@astro.auth.gr}

\date{\today}

\begin{abstract}
The most promising technique for the control of neoclassical tearing modes in tokamak experiments is the compensation of the missing bootstrap current with electron-cyclotron current drive. In this frame, the dynamics of magnetic islands has been studied extensively in terms of the modified Rutherford equation, including the presence of current drive, either analytically described or computed by numerical methods. In this article, a self-consistent model for the dynamic evolution of the magnetic island and the driven current is derived, which takes into account the island's magnetic topology and its effect on the current drive. The model combines the modified Rutherford equation with a ray-tracing approach to electron-cyclotron wave-propagation and absorption. Numerical results exhibit a decrease in the time required for complete stabilization with respect to the conventional computation (not taking into account the island geometry), which increases with increasing initial island size and radial misalignment of the deposition.
\end{abstract}


\maketitle

\section{Introduction}

An important issue for the ITER design, which is under careful investigation, is the stabilization of Neoclassical Tearing Modes (NTMs) by using Electron-Cyclotron Current Drive (ECCD). NTMs inhibit the optimal operation of tokamak devices because the generated magnetic islands, on surfaces with rational safety factor value $q=m/n$, taper the plasma energy and angular momentum leading to gradual loss of confinement and finally disruption \cite{Hegna97,Wilson08}. It is estimated that NTMs will be dynamically unstable in ITER, due to the high plasma pressure scenarios to be envisaged for achieving effective fusion results, and that both the $2/1$ and $3/2$ modes will be present \cite{LaHaye06}. The successful control of NTMs with ECCD has been demonstrated in large-scale experiments like AUG, DIII-D and JT-60U \cite{Zohm07,Maraschek12}, and these results have been forming the basis for the design of a corresponding control system for ITER.

There has been a lot of research on the properties of the NTM stabilization by EC waves, mainly investigating the effect of localized wave power deposition and current drive on the magnetic island growth (reviews on this topic are \cite{LaHaye06} and \cite{Urso09}). The evolution of ECCD-driven magnetic islands has been extensively analyzed in terms of the Modified Rutherford Equation (MRE), a modification of the Rutherford equation for classical tearing modes with the inclusion, among other physics, of the bootstrap current and the external current drive \cite{Zohm07,Rutherford73}. It has been understood in both theory and experiment that, in order to succeed in a more effective mode stabilization, the cyclotron resonance should be highly localized around the island's O-point and the direction of the driven current should be aligned with the equilibrium bootstrap current.

The common knowledge that NTM stabilization will probably be a major issue in ITER (and maybe also in DEMO) has led to an effort of improving the modeling of magnetic island dynamics in the presence of stabilizing ECCD, both on a general theoretical basis as well as in terms of simulations oriented to specific devices, so that the goal of successful validation with experiments can be reached. Many effects that may play a role in the stabilization effort and that were excluded from the earlier exploratory research, like e.g. local electron transport, diamagnetic rotation \cite{Ayten12}, wave-induced electric field (\cite{Rosa10}, with comments by \cite{Westerhof11}), edge turbulence \cite{Tsironis09} and EC beam misalignment, as well as the possible advantage of early ECCD application \cite{Lazaros07}, are currently being analyzed by using different techniques.

Regarding the modeling of ECCD, a variety of methods are available for computing the wave propagation, resonant absorption and driven current. For the propagation, one is mainly based on the asymptotic methods originating from geometric optics \cite{Bernstein75}. In ray tracing, canonical equations provide the position and the wavenumber along the ray trajectory in terms of the derivatives of the dispersion relation \cite{Friedland80}. The rays do not interact among themselves, therefore wave effects like diffraction are not properly accounted for. In quasi-optics, a beam is simulated as a set of interacting rays, and the basic wave effects are retained \cite{Mazzucato89}, whereas pWKB beam tracing is a more convenient description, based on a combination of ray tracing with a set of functions for the beam width and the wave-front curvature \cite{Pereverzev98}. The computation of the wave damping breaks down to evaluating the linear absorption coefficient along the ray path \cite{Westerhof89}, whereas the current drive may be calculated analytically with the linear adjoint method \cite{Cohen87}. There is a number of advanced codes implementing the above schemes, and the results in some cases are in sufficient agreement with the experiment \cite{Prater08}.

In almost all the codes that simulate ECCD-based NTM stabilization, the analysis of the wave evolution is done in the unperturbed magnetic configuration, assuming sufficient alignment of the EC resonance with the island's O-point on the flux surface of interest, yet ignoring effects from the island topology. This approach does not introduce an error in the computation of the ray propagation due to the smallness of the amplitude of the magnetic perturbation. However, islands bring up significant changes in the magnetic topology and the plasma profiles in comparison to the axisymmetric case: The different nesting of the flux surfaces and the flattening of the pressure profile within the island may play a crucial role in the wave deposition \cite{Isliker12}. Moreover, the ECCD efficiency in the presence of an island has been shown to be much different from the axisymmetric case, leading to different estimates for the minimum current required for stabilization \cite{Rosa10,Hamamatsu07}.

There are many efforts to introduce effects owed to the island geometry in the MRE formalism, since, in general, changes to the island shape are neglected by considering only the dominant harmonic of the perturbed flux. In this direction, the MRE has been reformulated to include a model for asymmetric island deformation \cite{Lazzaro09}, with the goal to ascertain the additional requirements that an ECCD-based NTM control system must satisfy if the magnetic islands undergo deformations induced e.g. by a sheared viscous flow. The results show that such deformations nonlinearly affect the time-scale of the island growth and can introduce a severe reduction in the ECCD control capability. This may have consequences for the localization of the beam around the O-point and the estimate of the minimum power needed for island quench.

A different modeling option for improving the accuracy in the description of the island topology within the frame of the MRE is to introduce a set of device-dependent parameters as multipliers of each term, and determine these by fitting the MRE solution to experimental results from specific devices \cite{Urso09,Urso05}. In this fashion, deviations owed to simplifications in the modeling (like the adoption of cylindrical geometry in some cases) are minimized. The main results indicate that in ITER, if the wave beam and the island's O-point are sufficiently aligned, the minimum wave power required to stabilize the $2/1$ and $3/2$ modes is always within the capabilities of the planned EC system. This suggests that the most challenging task for NTM control in ITER might be the optimization of the alignment between the ECCD injection and the island motion.

The effect of the island topology on ECCD is studied also in terms of electron transport models. Simulations of the $2/1$ NTM have been performed using the transport code TOPICS \cite{Isayama07}, combined with an experimentally-fitted version of the MRE, and the temporal evolution of the island width as observed in JT-60U was found to be well reproduced by the model. The simulation also showed that increasingly precise injection is required for smaller EC power, and that the allowable error in the ECCD location does not increase significantly even for large EC wave power. These results are similar to the ones from DIII-D for a $3/2$ NTM \cite{LaHaye02}. In addition, the TOPICS simulation predicts that the ECCD deposition width has a strong effect on the NTM control. This has experimentally been demonstrated in AUG, where it has been shown that narrow ECCD deposition could stabilize an NTM more effectively \cite{Maraschek12}.

The Monte-Carlo method has been applied in the study of the characteristics of ECCD in the magnetic island as a test-particle problem in the island topology, including Coulomb collisions and the EC quasilinear diffusion process \cite{Hamamatsu07}. The driven current was found to remain localized within the helical flux tube and its profile tended to have a peak around the O-point, whereas the ECCD efficiency was computed to be larger than in the axisymmetric case. The enhancement of the current occurs because the resonant electrons are well-confined in the smaller volumes defined by the island, despite the nonlinear effect introduced by  high power density \cite{Kamendje05}. In such cases, where the control is achieved by a current density driven around the O-point, the required power can be significantly reduced.

A self-consistent treatment of the wave-island interaction has been made with the numerical code NIMROD \cite{Sovinec04}, by augmenting the code with a quasilinear model for the basic EC wave physics to a closed set of RF-MHD equations \cite{Jenkins10}. The investigation of the effect of ECCD on the dynamical behavior of NTMs demonstrated the complete suppression of initially saturated $2/1$ and $3/2$ modes by the application of toroidally-symmetric ECCD, as well as the consequences of the shifting of the mode flux surface in response to the injected current and of the spatial ECCD misalignment were explored, effects which cannot be easily described by models based on the MRE. In a further development, the incorporation of the ability to use data from ray tracing codes in the NIMROD simulations, in order to determine the amplitude and spatial localization of the induced electromotive forces, has been theoretically established in terms of an advanced RF-MHD model \cite{Jenkins12}.

In this paper, a self-consistent computation of the dynamic evolution of the NTM growth in the presence of stabilizing ECCD is performed, on the basis of linear wave-particle physics and including the effect of the island geometry on ECCD, as presented in recent work \cite{Isliker12}. The connection of the effect of the island topology on the EC wave propagation and the resonant electron transport with the NTM dynamics, through the modification of the ECCD, is formulated by coupling the generalized Rutherford model with a ray-tracing solver. The geometric effect is introduced in this approach in terms of a fitting function that connects the island width and the driven current density, and which is determined with the ray-tracing code. Then, the modified Rutherford equation is solved as a function of time, with the instantaneously required values of the ECCD density being given through the mentioned and pre-computed fitting function.

The structure of the paper is as follows: In \Sref{model} the self-consistent numerical model is presented with a synopsis of the theory behind it, focusing on the coupling of the wave-field solution with the MRE, and in \Sref{results} the numerical results are presented and analyzed for the different cases studied. Finally, in \Sref{conclusion} the main results are summarized and the limitations of our model are discussed.

\section{Overview of the self-consistent model}
\label{model}

Keeping the focus on the effect of the island geometry on the ECCD deposition, the most important aspects to be considered are the influence of the helical magnetic field on the wave propagation and on the determination of the resonance region, the flattening of the radial profiles of the plasma electron density and temperature within the island region, and the structure of the volumes of the perturbed flux surfaces into which the wave power is deposited. The simulation tool we use for the computation of the EC propagation in a magnetic configuration that includes islands is the ray-tracing code CODERAY \cite{Isliker12}, whereas the connection of the results for the wave to the dynamics of the NTM suppression is made by solving numerically the modified Rutherford equation.

In the ray tracing asymptotic technique, which stems from geometric optics theory, the propagation of waves is  formulated in terms of a set of ordinary differential equations (ODEs), which may be integrated by means of a standard numerical solver and therefore are simpler to tackle than the (partial differential) Helmholtz equation. The canonical equations for the ray propagation in this framework are \cite{Friedland80,Stix92}
\begin{subequations}
\label{rayeqstime}
\begin{eqnarray}
\frac{d\mathbf{k}}{dt}=&& -\frac{\partial\omega}{\partial\mathbf{r}},\\
\frac{d\mathbf{r}}{dt}=&& \frac{\partial\omega}{\partial \mathbf{k}},
\end{eqnarray}
\end{subequations}
where the ray position $\mathbf{r}$ and the wave vector $\mathbf{k}$ are canonical variables, and the wave frequency $\omega$ plays the role of the Hamiltonian. The solution of \eref{rayeqstime} determines the propagation in the plasma as a function of time, with characteristic time-scale the wave period. A new independent variable $\tau$ can be introduced such that the equations take a form where $t$ is replaced by $\tau$ and $\omega$ by a new Hamiltonian $\mathcal{H}$ that  is the solvability condition of the dispersion relation viewed as a function of $(\mathbf{r}$, $\mathbf{k})$. With $\mathbf{\epsilon^{h}}$ the Hermitian part of the plasma dielectric tensor and $c$ the light speed in vacuum, we have
\begin{equation}
\label{hamray}
\mathcal{H}=\det\left [\left (\frac{\omega}{c}\right )^{2}\left (-k^{2}\mathbf{I}+
\mathbf{k}\mathbf{k}\right )+\mathbf{\epsilon^{h}}\right ].
\end{equation}
This formalism is most appropriate in the case of monochromatic wave propagation in stationary plasma, because it provides a relation of the integration step with the time which is parametric-dependent only on the frequency, and therefore the time dependence can be further neglected. Assuming cold plasma propagation, one may adopt the cold plasma dielectric tensor \cite{Stix92}, which gives the final expression for the Hamiltonian used here (for more information on the wave code, see \cite{Isliker12}).

In the context of geometric optics, the wave propagation is a zero-order process, whereas the absorption, and consequently the driven current, are described by the first-order equations. The wave absorption is computed along the ray path in terms of the imaginary part of the wave vector, as determined from the dispersion relation \cite{Westerhof89}. The evolution of the absorbed wave power $P_{abs}$ is given then by
\begin{equation}
\label{powdamp}
\frac{dP_{abs}}{d\tau}=-2\mathrm{Im}({\mathbf{k}})\cdot\mathbf{v_g}(P_0-P_{abs})   ,
\end{equation}
where $\mathbf{v_g}$ is the group velocity and $P_0$ the injected wave power. With the absorbed power along the ray path known, the power $dP_{abs}$ deposited in a small radial interval can be calculated from \eref{powdamp}, and a division by the volume $dV_{abs}$ contained between the two flux surfaces enclosing the radial interval gives the absorbed power density. The total driven current $I_{CD}$ over the absorbed power defines the current drive efficiency $\zeta_{CD}$ ($r_{maj}$ is the major plasma radius)
\begin{equation}
\label{cdeff}
\zeta_{CD}=2\pi r_{maj}\frac{I_{CD}}{P_{abs}}.
\end{equation}
Following \cite{Cohen87}, the ECCD efficiency is computed in terms of the linear adjoint method, based on a Green's function formulation with the magnetic field approximated as a square well in order to obtain an analytic solution, and it includes the effects of trapped particles, ion-electron collisions and the spatial variation of the collision operator.

As an input to the wave solver, 
the magnetic field topology and the radial
profiles of the electron density/temperature of the plasma must be provided. The non-axisymmetric magnetic configuration used here has been formulated as in \cite{Isliker12}. To start with, the total magnetic field is expressed as
\begin{equation}
\label{mftotisl}
{\bf B}=\frac{1}{r}\frac{\partial\psi_t}{\partial r}\hat{\mathbf{e}_{r}}+\frac{1}{R}\frac{\partial\psi_p}{\partial r} \hat{\mathbf{e}_{\theta}}-\frac{1}{rR}\left (\frac{\partial\psi_p}{\partial\theta}+ \frac{\partial\psi_t}{\partial\varphi}\right )\hat{\mathbf{e}_{\phi}},
\end{equation}
with $\psi_t$, $\psi_p$ the toroidal and poloidal flux functions, $r$, $\theta$, $\varphi$ the radial, toroidal and poloidal coordinates with unit base vectors $\hat{\mathbf{e}_{r}}$, $\hat{\mathbf{e}_{\theta}}$, $\hat{\mathbf{e}_{\varphi}}$, and with $R=r_{maj}+r\cos\theta$. For the axisymmetric part of the magnetic field, corresponding to the background equilibrium, the expression used is the one known as "vacuum magnetic field"
\begin{subequations}
\label{stmagfield}
\begin{eqnarray}
B_{t0}(r,\theta)=&&\frac{B_{0}}{1+\epsilon_A (r)\cos\theta},\\
B_{p0}(r,\theta)=&&\frac{\epsilon_A (r)}{q\left(r \right)}B_{t0}(r),
\end{eqnarray}
\end{subequations}
where $B_{t0}$, $B_{p0}$ are the toroidal and poloidal component, respectively, $B_{0}$ the toroidal field on the magnetic axis, $\epsilon_A(r)=r/R_{0}$ the inverse aspect ratio and $q(r)=d\phi/d\theta$ is the safety factor, chosen as a monotonically
increasing rational function, see \cite{Isliker12}. The flux functions $\psi_{t0}$, $\psi_{p0}$ corresponding to the fields can be calculated from the relations $\partial_r \psi_{t0}=rB_{t0}$ and $\partial_r\psi_{p0}=RB_{p0}$. The local magnetic field structure of the island is described by
a perturbation $\psi_{p1}$ to the  poloidal flux, $\psi_p=\psi_{p0}+\psi_{p1}$,
with
\begin{equation}
\label{pertmagfield}
\psi_{p1}(r,\theta,\phi) = \varepsilon_{mn}(r)\cos(m\theta-n\phi),
\end{equation}
see e.g.\ \cite{Hazeltine92}, where $\varepsilon_{mn}$ is the perturbation strength and $m,n$ are the mode numbers of the NTM. For an implementation of the NTM topology, one has to specify
$\varepsilon_{mn}(r)$,
and, as described in detail in \cite{Isliker12},
we use the low-order approximation of  \cite{Ren98}
to the self-consistent expression given in  \cite{Militello06},
\begin{equation}
\psi_{p1} = - \frac{r}{m} \varepsilon_{mn}^{(0)}\left(1+\frac{r-r_s}{\alpha_\pm}\right)\cos(m\theta-n\phi),
\end{equation}
with $\alpha_\pm$ the slopes, and $r_s$ the radius of the resonant surface,
determined through the equation $q(r_s)=m/n$.

Apart from the changes in the magnetic topology, an excited NTM causes the plasma pressure to assume a constant value inside the separatrix of the island chain. This flattening in the pressure profile leads in turn to a flattening in the electron density and temperature profiles. In order to model this alteration of the plasma profiles, we assume the density and temperature profiles to be parabolic functions outside the island, as in the unperturbed case, to be constant within the island region and to equal to the density and temperature values at the outer island boundary, and to be continuous at the inner island boundary (see \cite{Isliker12} for details).

For the computation of the power absorption in the presence of the island, the calculation of the plasma volume between two adjacent flux surfaces is required. Again following \cite{Isliker12}, the total volume $V_{abs}$ contained inside a flux surface is
\begin{equation}
\label{islvol}
V_{abs}=-\frac{1}{n}\int_0^{2\pi n}\int_{\xi_1}^{\xi_2}\int_{r_1}^{r_2}\left (r_{maj}+r\cos\theta\right )\, r dr \,d\xi \, d\theta,
\end{equation}
with $\xi=m\theta-n\phi$ the helical angle in the direction transverse to the line through the island's O-point, and $r_1$, $r_2$, $\xi_1$, $\xi_2$ the integration limits. The definition of the integration limits, as given in \cite{Isliker12}, requires an analytical labeling $\Omega$ of the flux surfaces in the island region. This expression has been derived in \cite{Isliker12} as
\begin{equation}
\Omega =
 \frac{1}{2}   (r-r_s)^2
 +\frac{r}{r_s} \Omega_s
 \left(1+\frac{r-r_s}{\alpha_{\pm}}\right)
\cos\left(\xi \right),
\label{omegan}
\end{equation}
with
\begin{equation}
\Omega_s = \frac{r_s \varepsilon_{mn}^{(0)}}{m\, r_{maj} \,
\left(\partial_{rr} \psi_{p0}\right)\Big |_{r_s}}
\end{equation}
the value of $\Omega$ on the separatrix,
and the island half-width $W_{1/2}$ approximately is
\begin{equation}
 W_{1/2} =  \sqrt{\frac{ 2 r_s \varepsilon_{mn}^{(0)}}{m\, r_{maj} \,
\left(\partial_{rr} \psi_{p0}\right)\Big |_{r_s}} }.
\label{iwidth}
\end{equation}
Having these at hand, and the ray-tracing data, the absorbed power per unit volume can be evaluated as
\begin{equation}
\label{pabsvol}
\frac{dP_{abs}}{dV_{abs}}=\frac{dP_{abs}}{d\tau}\left (\frac{dV_{abs}}{d\tau}\right )^{-1}.
\end{equation}

The established model for the dynamic evolution of the NTM is the modified Rutherford equation, which is based on a generalization of the classical Rutherford equation for tearing modes \cite{LaHaye06,Rutherford73}. In the case of classical tearing modes, only the Ohmic current contributes, whereas for NTMs other currents that flow in the island region, as appearing in neoclassical transport, need to be accounted for, the most important of which is the bootstrap current. In this context, and with the inclusion of the stabilizing ECCD, the MRE becomes \cite{Urso09}
\begin{equation}
\label{mre}
\frac{\tau_r}{r_s}\frac{dW}{dt}=r_s\Delta'_{\beta}=r_s\left (\Delta'+\Delta'_{BS}+\Delta'_{CD}\right),
\end{equation}
where $W$ is the full island width, $\tau_r=0.82\mu_0r_s^2/\eta_p$ the resistive time-scale of the plasma ($\eta_p$ the plasma resistivity), $\Delta'$ the neoclassical stability index, including the classical index and a term connected to the nonlinear island saturation, $\Delta'_{BS}$ the stability index corresponding to the bootstrap current, and $\Delta'_{CD}$ the term which represents the stabilizing effect of the driven current. These stability indices are given as \cite{Hegna97,Urso09}
\begin{subequations}
\label{mredeltas}
\begin{eqnarray}
&& \Delta'=-\frac{m}{r_s}-\frac{W}{2.44r_{min}^2}, \\
&& \Delta'_{BS}=\frac{\sqrt{\epsilon_A}\beta_P}{W}\frac{L_q}{L_p}\left (\frac{W^2}{W^2+W_d^2}+ \frac{W^2}{W^2+28W_b^2}-\frac{W_{pol}^2}{W^2}\right ), \\
&& \Delta'_{CD}=-\frac{16}{\pi}\frac{\mu_0L_q}{B_\theta}\frac{1}{W^2}I_{CD}\eta_{CD}U(t-t^{on}_{CD}).
\end{eqnarray}
\end{subequations}
In the above, $r_{min}$ is the minor radius, $\epsilon_A=r_{min}/r_{maj}$ the aspect ratio of the tokamak, $\beta_P$  the ratio of the plasma and magnetic pressures, $L_q$ and $L_p$ the shear lengths of the safety factor and the plasma pressure, $t^{on}_{CD}$ the time when ECCD is turned on (with $U$ the Heaviside step function \cite{Arfken05}), $I_{CD}$ is the driven current ($\eta_{CD}$ is defined below), and $W_d$ ,$W_b$, $W_{pol}$ are characteristic threshold values for the island width: $W_d$ is the critical width for classical destabilization, $W_b$ the width below which banana orbits contribute significantly to the bootstrap current, and $W_{pol}$ the width below which the current generated in response to the diamagnetic island rotation is significant (for more details on the physics of the different parameters and terms see \cite{Hegna97,LaHaye06,Mikhailovskii03} and references therein).

In the computation of the term $\Delta_{CD}^{'}$, the total driven current is expressed in terms of the current density as $I_{CD}=\pi^{3/2}r_sd_{CD}j_{CD}$, where $j_{CD}$ is assumed to have a Gaussian radial profile around the O-point with peak value $j_{CD0}$ and width $d_{CD}$
\begin{equation}
\label{cddens}
j_{CD}=j_{CD0}\exp{\left [-\frac{\left (r-r_s-r_{mis}\right )^2}{d_{CD}^2}\right ]},
\end{equation}
and $r_{mis}$ denotes a small distance of radial misalignment between the deposition center and the O-point. The efficiency of the ECCD injection in stabilizing the NTM is described by the factor $\eta_{CD}$ appearing in \eref{mredeltas}. $\eta_{CD}$ is a measure of the geometrical optimization of the deposition on the basis of the ECCD radial profile width as compared to the island width (not to be confused with the current-drive efficiency $\zeta_{CD}$, given in \eref{cdeff}). This factor, among other things, depends heavily on the synchronization of the island motion with the wave power constancy or modulation. For locked islands, stabilization is possible only if the O-point position is geometrically accessible to the EC system and no power modulation is required. When the islands rotate, conventionally $\Delta'_{CD}$ should be obtained by averaging \eref{cddens} over the rotation period, in order to assess properly the distribution of the wave power over the different island phases. In our model, we have assumed that (a) for locked modes, the EC deposition around the O-point is feasible, and (b) for rotating modes, the power is modulated exactly at the island rotation frequency and the power-on phase is exactly centered around the O-point passage through the beam. In this framework, $\eta_{CD}$ has the form \cite{Urso05}
\begin{equation}
\label{etaeccd}
\hspace{-2.5cm}\eta_{CD}\hspace{-0.1cm}=\hspace{-0.1cm}0.07\hspace{-0.1cm}\left (\hspace{-0.1cm}\frac{W}{d_{CD}} \hspace{-0.1cm}\right )^{\hspace{-0.1cm}2}\hspace{-0.1cm}+\hspace{-0.1cm}\left [0.34\hspace{-0.1cm}-\hspace{-0.1cm} 0.07\hspace{-0.1cm}\left (\hspace{-0.1cm}\frac{W}{d_{CD}}\hspace{-0.1cm}\right )^{\hspace{-0.1cm}2}\right ]\hspace{-0.2cm}\left [\frac{0.3W}{d_{CD}}\hspace{0.05cm} U\hspace{-0.1cm}\left (\hspace{-0.1cm}2\hspace{-0.1cm}- \hspace{-0.1cm}\frac{W}{d_{CD}}\hspace{-0.1cm}\right )\hspace{-0.1cm}+\hspace{-0.1cm}\exp{\hspace{-0.1cm}\left (\hspace{-0.1cm}-\frac{d_{CD}}{W}\right )}U\hspace{-0.1cm}\left (\hspace{-0.1cm}\frac{W}{d_{CD}}\hspace{-0.1cm}- \hspace{-0.1cm}2\hspace{-0.1cm}\right )\hspace{-0.1cm}\right ].
\end{equation}

In including the current drive modification caused by the island into the NTM dynamics self-consistently, the direct coupling of the MRE with the ray tracing algorithm based on \eref{rayeqstime}, with time as the common independent variable, exhibits the problem of the vastly different time-scales of evolution: In ITER, the wave period $T$ is of the order of at most $10^{-10}$ s, whereas the resistive time-scale is of the order of at least $10^{-4}$ s. It becomes obvious that the island evolution, being a "slow" process, will not be affected by wave effects on a time-scale comparable to $T$, whereas the "fast" wave propagation could potentially also be affected on a time-scale of the order of $t_r$. Since it would be obligatory, in order to obtain a physically consistent solution, to time-step the problem on the slow time-scale, an inefficiency in the computational scheme would result.

In order to reduce the computational burden for treating the problem, we evolve the plasma instability process
on the $t_r$ time-scale during the EC wave propagation. For better efficiency, we actually progress only the MRE and, at each time step (which is comparable to $t_r$), we compute the radial profile of the ECCD with the ray tracing code, using the instantaneous value of the island width for determining the magnetic field perturbation and other related parameters. The resulting new value of the driven current density, which includes now the modification caused by the change of the island width, is then provided back to the MRE in order to compute the next step that yields the new island width, and so on. Non-strictly speaking, this defines a self-consistent model for the evolution of the magnetic island width and the ECCD profile.

One may further disengage from running the wave code at each time-step of the MRE: Since the problem setup for the wave evolution is linear, as the propagation/absorption is treated in terms of a linear dielectric response tensor and the current drive is computed with the linear adjoint method, what is practically needed for including the effect of the island on the ECCD is an analytic or tabular function connecting $W$ and $j_{CD}$. According to \eref{mredeltas} and \eref{etaeccd}, in order to express the relation $j_{CD}=f(W)$ for a specific injection setup ($P_0$,$\theta_l$,$\phi_l$), one needs to determine, via the ray-tracing data, the dependence of $j_{CD0}$ and $d_{CD}$ on $W$ ($r_s$ depends only on the $q$-profile). In this sense, one can use the wave code to compute the parameters of the driven current for many different $W$ values in order to build a table of the corresponding $j_{CD0}$ values, and then, in evolving the MRE, at a certain time $t$ when the island width is $W$, one computes the required value of $j_{CD0}$ by linear interpolation/extrapolation of the tabulated values.

\section{Numerical results}
\label{results}

This section contains the numerical results for the estimation of the effect of the magnetic topology on the NTM dynamics via the alteration of the driven current, using the self-consistent model previously described. As described above, we incorporate the island geometry via a set of tabulated values connecting the island width and the driven current density, which are computed by the ray-tracing code CODERAY. Thereafter, the modified Rutherford equation is solved numerically, with the required values of the ECCD being determined from the pre-computed and tabulated values of $j_{CD0}$ vs $W$.

Regarding the characteristics of the stabilization process, especially its global efficiency and speed in both cases of locked and rotating islands, the important effects to be investigated here have to do with the following parameters: (a) the initial value of the island width, (b) the radial misalignment of the ECCD peak with respect to the O-point, and (c) the specific time instant at which the wave power is turned on.

\subsection{Ray tracing computations of ECCD}
\label{rtcomp}

In the ray tracing computations, we use a magnetic equilibrium with islands, as generated by an NTM of order $3/2$, in combination with varying magnetic perturbation strengths (and thus island widths), and we also include the effect of the flattening of the plasma pressure profile (for details see \cite{Isliker12}). The wave is launched from the outermost flux surface ($r=r_{min}$), at a poloidal angle such that the ray propagation targets extremely close to the O-point (the necessity for $r_{mis}\approx 0$ will be analyzed later on) and a toroidal magnetic field for which the EC resonance layer is located around the O-point. The plasma and wave parameters are the ones foreseen in ITER: The major and minor radii are $r_{maj}=6.2$ m and $r_{min}=1.9$ m, the  magnetic field on the tokamak magnetic axis is $B_0=5.51$ T, the electron density and temperature follow parabolic profiles with values at the plasma center equal to $n_e(0)=10^{20}$ m$^{-3}$,  $T_e(0)=10$ KeV and at the edge $n_e(r_{min})=10^{19}$ m$^{-3}$, $T_e(r_{min})=1$ KeV, the $q$-profile is also parabolic with $q(0)=1$, $q(r_{min})=4$ (see \cite{Isliker12}), the wave frequency is $\omega/2\pi=170$ GHz (fundamental O-mode) and the initial wave power is $P_0=10$ MW.

\begin{figure}
\includegraphics[scale=0.8]{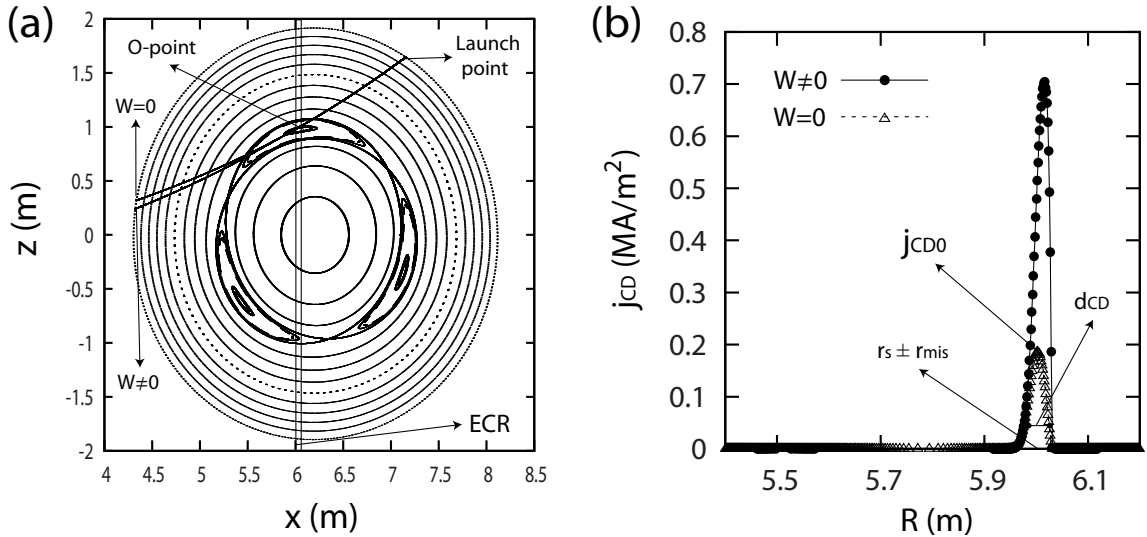}
\caption{EC wave propagation and current drive, as computed by the ray tracing code CODERAY, for the case of a $m/n=3/2$ mode in ITER with $W_0=30$ cm (in comparison with the same computation performed in axisymmetric geometry): (a) Poloidal projection of the ray propagation path, (b) radial profile of the ECCD density.}
\label{fig1}
\end{figure}

An indicative result from the wave code is shown in \Fref{fig1}, where, apart from the parameters mentioned above, the magnetic perturbation strength has a value such that the island width is $W_0=30$ cm. In \Fref{fig1}(a), we show the projection of the ray path onto the poloidal plane, in the presence of the NTM, and also compared to the corresponding path in the unperturbed equilibrium (i.e.\ in the absence of the mode). The injection angles of the EC wave beam are $\theta_l=-30^o$ poloidally and $\phi_l=-5^o$ toroidally. For the assumed profiles of $\mathbf{B}$, $q$, $n_e$, $T_e$, and the chosen value of $\omega$, the layer of the EC resonance is located around $R=6$ m, as marked in \Fref{fig1}(a) with the narrow region between the two vertical lines. In \Fref{fig1}(b), the radial profile of the ECCD density is visualized for the two cases, with and without an NTM present, respectively, and here we just note that the obvious characteristic differences and  effects have been analyzed and discussed in detail in \cite{Isliker12}.

The shape of the radial profile in \Fref{fig1}(b) highly resembles a Gaussian function curve, which implies that fitting the profile data against the Gaussian function in \Eref{cddens} is appropriate.
This fitting, provided that $r_{mis}\approx 0$ (as ensured during the ray tracing computations), yields the parameters $j_{CD0}$, $d_{CD}$ for a given profile and value of $W$, and it has been repeated for varying values of $W$. The resulting data for the peak value and the $1/e$-width of the ECCD profile, as a function of the island width or the  dimensionless magnetic perturbation amplitude, are given in \Tref{tab1}. The only parameter that has changed with respect to above is $\phi_l=0^o$, and in the first row the results for the unperturbed case $W=0$ is included, in order to perform comparisons. The physical reason for the dependence of $j_{cd0}$ and $d_{cd}$ on $W$ stems from the fact that, in the presence of the island,  the wave power is deposited into volumes smaller than those in its absence, which in turn leads to larger values of the absorbed power density and the driven current. As the island width increases, so is the magnetic perturbation strength, and the flux surface nesting in the interior of the island becomes more complicated, therefore this effect appears stronger. The dependence of $d_{cd}$ on $W$ is much weaker (the overall increase is 0.16 cm over a 20 cm increase of the island size), however it was included in the modeling for consistency reasons.

\begin{table}
\caption{The numerical values of $j_{CD0}$ and $d_{CD}$, for different values of $W$ (or $\epsilon_{32}$), determined through a Gaussian fit to the ray-tracing data (ECCD profiles).}
\vspace{0.4cm}
\begin{indented}
\item[]
\begin{tabular}{llllllllll}
\br
$\epsilon_{32}$ &&& $W$ (m) &&& $j_{CD0}$ (MA/m$^2$) &&& $d_{CD}$ (m) \\
\mr
0.000 &&& 0.0000 &&& 0.1102 &&& 0.01904 \\
0.012 &&& 0.1134 &&& 0.2123 &&& 0.01984 \\
0.015 &&& 0.1247 &&& 0.2216 &&& 0.01985 \\
0.017 &&& 0.1350 &&& 0.2283 &&& 0.01991 \\
0.020 &&& 0.1450 &&& 0.2368 &&& 0.02002 \\
0.023 &&& 0.1553 &&& 0.2451 &&& 0.02016 \\
0.026 &&& 0.1654 &&& 0.2528 &&& 0.02023 \\
0.029 &&& 0.1749 &&& 0.2614 &&& 0.02033 \\
0.032 &&& 0.1853 &&& 0.2732 &&& 0.02047 \\
0.036 &&& 0.1952 &&& 0.2864 &&& 0.02058 \\
0.039 &&& 0.2047 &&& 0.2986 &&& 0.02066 \\
\mr
\end{tabular}
\end{indented}
\label{tab1}
\end{table}

\begin{figure}
\vspace{0.1cm}
\includegraphics[scale=0.94,trim=0 0.35cm 0 0]{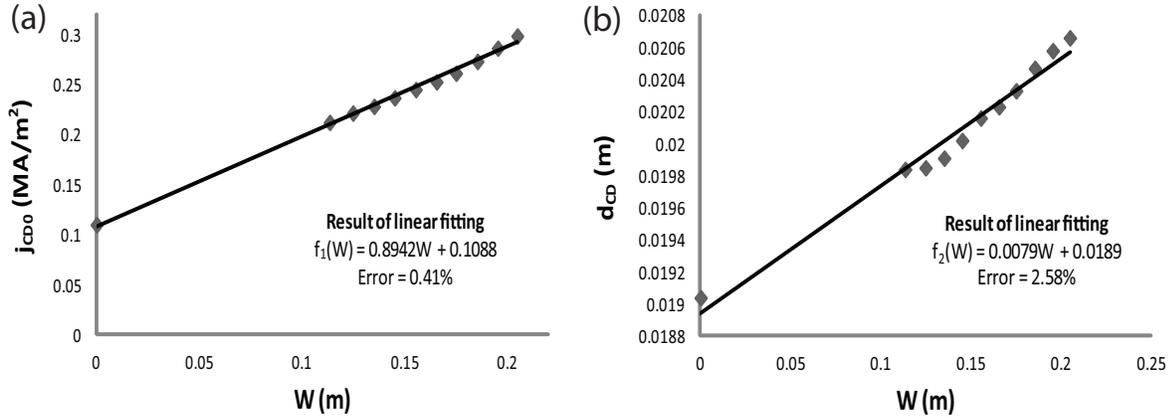}
\caption{Results of the linear fitting (least-squares method) to the tabulated values of (a) $j_{CD0}$ and (b) $d_{CD}$ in \Tref{tab1}, as a function of the island width.}
\label{fig2}
\end{figure}

For practical reasons, instead of interpolating/extrapolating $j_{CD0}$ and $d_{CD}$ from the values of \Tref{tab1} to arbitrary values of $W$, one may introduce appropriate functional forms $j_{CD0}=f_1(W)$ and $d_{CD}=f_2(W)$ and make them  specific by fitting them to the tabulated values. Since the ECCD computation is done in terms of the linear adjoint method, it is expected that the scaling of $j_{CD0}$ and $d_{CD}$ with $W$ will be, in a good approximation, linear. And indeed, as shown in \Fref{fig2}, the result of linear fitting is very successful in both cases: The regression error for $f_1(W)$ is less than $1\%$ and for $f_2(W)$ it is less than $3\%$. So, in the frame of the MRE evolution, we will use the fitted functions $f_1$ and $f_2$.

\subsection{Solution of the modified Rutherford equation}
\label{mresol}

In this section, the modified Rutherford equation is solved self-consistently by using the results on the current drive from the previous section, and the results are analyzed in comparison to those from the non-self-consistent case in axisymmetric geometry. The parameters used here are the same as in the wave computations, and the parameters specific to the MRE are chosen as $\beta_P=0.5$, $L_q=L_p=1$, $W_d=0.01r_s$, $W_b=0.02r_s$ and $W_{pol}=0.015r_s$. The radius of the flux surface where the mode resides is found by solving the algebraic equation $q(r_s)=3/2$, and the ECCD density (peak value, width and misalignment) are given by the fitted functions $f_1$ and $f_2$ from the previous section. Concerning the results, we mostly focus on the time-domain signal ($W$ vs $t$) and the phase diagram ($r_s\Delta'_\beta$ vs $W/r_{min}$).

\begin{figure}
\includegraphics[scale=0.83]{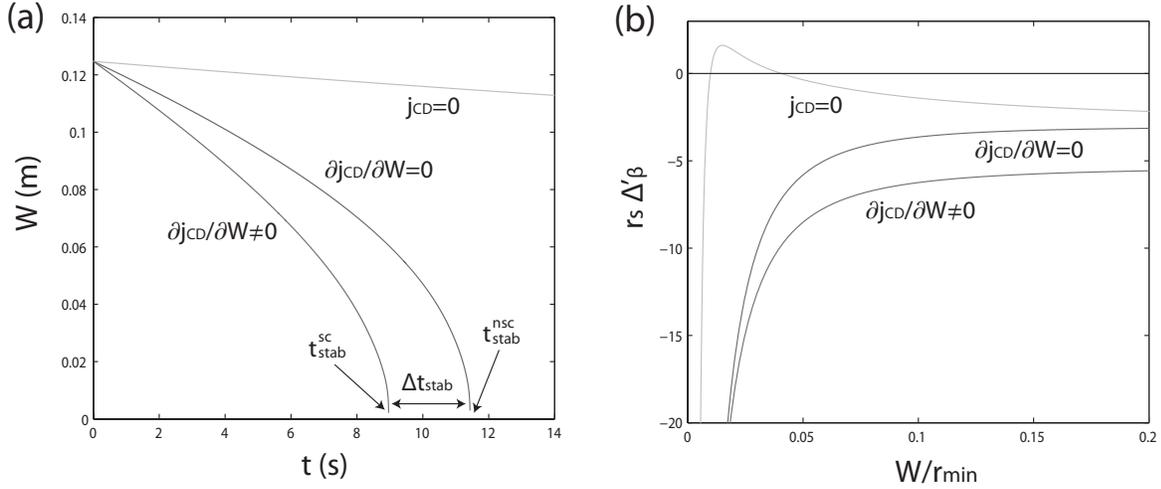}
\caption{Self-consistent solution of the modified Rutherford equation for the case of a $3/2$ mode in ITER with $W_0=12.47$ cm, in comparison with two cases of
the non-self-consistent computation in axisymmetric geometry, namely in the presence and absence of ECCD, respectively: (a) Time-evolution of the island width, (b) phase diagram of the MRE.}
\label{fig3}
\end{figure}

In \Fref{fig3}, the solution of the MRE for initial width $W_0=12.47\,$cm and $t^{on}_{CD}=0\,$s, $r_{mis}=1\,$cm for the EC power is presented, in the cases of no ECCD applied $(j_{CD}=0)$, of ECCD applied and computed in the axisymmetric geometry ($\partial j_{CD}/\partial W=0$) and of ECCD applied and computed self-consistently in the perturbed geometry ($\partial j_{CD}/\partial W\ne 0$). The evolution of $W(t)$, as seen in \Fref{fig3}(a), reveals that the NTM is stabilized faster on the basis of the self-consistent computation. This occurs because the ECCD density, as computed in terms of the self-consistent model, is always larger than the one in the axisymmetric case, due to the geometric effect of smaller flux-surface volumes (see \cite{Isliker12}), which ultimately leads to an enhancement of $\Delta^{'}_{CD}$. This can also be seen in \Fref{fig3}(b), which is actually an imprint of the growth dynamics, and where the phase curve from the self-consistent model attains larger negative values than in the other two cases.

In the following, we will further investigate the deviation appearing in the computation of the time required for complete stabilization between the two models. As "stabilization time" we define the time interval from the time-instant the mode is affected by the EC control system till the nullification of the island width, which we denote by $t_{stab}^{nsc}$ for the non-self-consistent computation and by $t_{stab}^{sc}$ for the self-consistent one. The deviation may then be defined as follows
\begin{equation}
\label{dtstablag}
\Delta t _{stab}=t_{stab}^{nsc}-t_{stab}^{sc},
\end{equation}
as illustrated also in the schematic representation in \Fref{fig3}(a). The parameters expected to affect the form of the self-consistent solution and the deviation from the standard result in axisymmetric geometry are the initial width $W_0$, the time-instant $t^{on}_{CD}$ and the maximum misalignment $r_{mis}$. In order to ascertain the effect of these parameters on the overall process, $t_{stab}^{nsc}$, $t_{stab}^{sc}$ and $\Delta t _{stab}$ have been computed for several values of $W_0$ and different combinations of $t^{on}_{CD}$, $r_{mis}$. The results of these computations are shown in \Tref{tab2}.
\begin{table}
\caption{Numerical values of the non-self-consistent and self-consistent computations of the stabilization time $t_{stab}^{nsc}$, $t_{stab}^{sc}$ and their deviation $\Delta t _{stab}$ in the case of a $3/2$ NTM in ITER for different values of the initial width $W_0$ and for different combinations of $t^{on}_{CD}$ and $r_{mis}$.}
\vspace{0.4cm}
\begin{indented}
\item[]
\hspace{-1.3cm}
\begin{tabular}{@{}lllllllllllllllllllllllll@{}}
\mr
$W_0$ (m) && $t_{stab}^{nsc}-t_{stab}^{sc}=\Delta t _{stab}$ (s) \\
\mr
$0.1134$ & $9.436-5.978=3.458$ & $12.359-7.766=4.593$ & $27.798-17.124=10.674$ \\
         & $11.518-7.923=3.596$ & $14.505-9.897=4.608$ & $29.496-19.165=10.331$ \\
$0.1247$ & $11.432-7.067=4.365$  & $14.866-9.138=5.728$  & $32.410-19.751=12.659$ \\
         & $13.531-8.997=4.534$  & $16.905-11.051=5.854$  & $34.045-21.866=12.179$ \\
$0.1350$ & $13.369-8.107=5.262$  & $17.273-10.442=6.831$  & $36.622-22.167=14.455$ \\
         & $15.402-10.032=5.370$  & $19.349-12.579=6.770$  & $38.099-24.178=13.921$ \\
$0.1450$ & $15.348-9.153=6.195$  & $19.704-11.737=7.967$  & $40.709-24.534=16.175$ \\
         & $17.495-11.038=6.457$  & $21.635-13.848=7.787$  & $42.046-26.425=15.621$ \\
$0.1553$ & $17.482-10.272=7.210$  & $22.291-13.109=9.182$  & $44.903-26.955=17.948$ \\
         & $19.485-12.386=7.099$  & $24.091-15.192=8.889$  & $46.124-28.770=17.354$ \\
$0.1654$ & $19.640-11.393=8.247$  & $24.901-14.481=10.420$ & $48.983-29.338=19.645$ \\
         & $21.620-13.518=8.102$  & $26.649-16.570=10.079$ & $50.128-31.139=18.989$ \\
$0.1749$ & $21.739-12.467=9.272$  & $27.401-15.799=11.602$ & $52.791-31.586=21.205$ \\
         & $23.605-14.580=9.025$  & $29.048-17.864=11.184$ & $53.843-33.311=20.532$ \\
$0.1853$ & $24.096-13.671=10.425$ & $30.188-17.261=12.927$ & $56.914-34.019=22.895$ \\
         & $25.855-15.766=10.089$ & $31.740-19.463=12.277$ & $57.884-35.692=22.192$ \\
$0.1952$ & $26.390-14.840=11.550$ & $32.884-18.673=14.211$ & $60.803-36.334=24.469$ \\
         & $28.062-16.921=11.141$ & $34.341-20.807=13.534$ & $61.718-37.956=23.762$ \\
$0.2047$ & $28.634-15.977=12.657$ & $35.497-20.036=15.461$ & $64.498-38.550=25.948$ \\
         & $30.234-18.038=12.196$ & $36.884-22.117=14.767$ & $65.342-40.134=25.208$ \\
\mr
$r_{mis}$ (m) & $0.00$ & $0.01$ & $0.02$ \\
$t_{CD}^{on}$ (s) & $0$ (1$^{st}$ line for each $W_0$) & $2$ (2$^{nd}$ line for each $W_0$) \\
\mr
\end{tabular}
\end{indented}
\label{tab2}
\end{table}

The concept of the stabilization time and its behavior as a function of the ECCD have been analyzed in previous studies (see e.g. \cite{Maraschek12,Urso09}), and since our work emphasizes the difference appearing in the estimation of the stabilization time when the effect of the island topology is taken into account self-consistently, an extensive analysis of stabilization times will not be made here. Just to mention, as seen in \Tref{tab2}, the values of $t_{stab}$ range from $9\,$s to $65\,$s, depending mainly on $W_0$ and $r_{mis}$. The stabilization times computed are nearly the same as the ones presented in \cite{Urso09}, where complete stabilization was found to occur roughly in a minute for the 3/2 NTM, and less than the characteristic NTM growth time expected in ITER (around 100 s). In the modeling set-up, the wave power may safely be considered to be active for all this time interval, since a realistic high-power ECRH pulse from the 1 MW gyrotron planned for ITER can last up to $400\,$s.

\begin{figure}
\includegraphics[scale=0.8,trim=0 0.25cm 0 0]{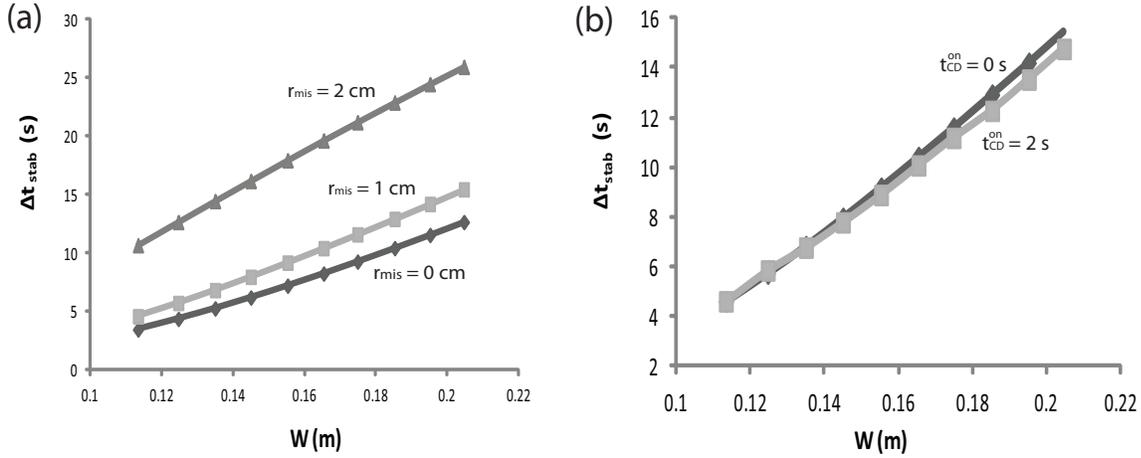}
\caption{Dependence of the scaling of the time-lag $\Delta t_{stab}$ with $W_0$ on $t^{on}_{CD}$ and $r_{mis}$: Diagram of $\Delta t_{stab}$ vs $W_0$ for (a) $t^{on}_{CD}=0$ s and three different values of $r_{mis}$, (b) $r_{mis}=0.01$ m and two different values of $t^{on}_{CD}$.}
\label{fig4}
\end{figure}

For each one of the six combinations of $r_{mis}$ and $t^{on}_{CD}$, the time-lag $\Delta t _{stab}$ is an increasing function of $W_0$, meaning that for initially larger islands the self-consistent model predicts a faster stabilization of the mode. This is visualized in \Fref{fig4}(a), where $\Delta t_{stab}$ is plotted against $W_0$ for $t^{on}_{CD}=0$ s and three different values of $r_{mis}$. The specific scaling occurs because the geometric effect on $\Delta^{'}_{CD}$, which results in the increase of the latter, becomes more important when the size of the island is larger. Furthermore, after a closer examination of \Tref{tab2}, the first impression is that the scaling of $\Delta t_{stab}$ with $W_0$ does not depend on $t^{on}_{CD}$. The respective plot of $\Delta t_{stab}$ vs $W_0$ is shown in \Fref{fig4}(b), for $r_{mis}=0.01$ m and two different values of $t^{on}_{CD}$. As a matter of fact, the time-instant $t^{on}_{CD}$ just determines the island width in the beginning of the stabilization effort, which, for the values of $t^{on}_{CD}$ occurring in the experiments ($<2$ s) and the slow evolution of the island when the ECCD is off (see \Fref{fig3}), retains a value very close to $W_0$.

On the contrary to the above, there is clearly a dependence of the scaling of $\Delta t_{stab}$ with $W_0$ on $r_{mis}$. In \Fref{fig4}(a), this appears in the form of a parametric up-shift of the scaling relation. This effect can be expected, because as $r_{mis}$ increases the quantity of deposited ECCD in the island region decreases and, whereas $\Delta^{'}_{CD}$ decreases within each computation alone, the difference of $\Delta^{'}_{CD}$ values coming from the self-consistent model and the standard computation continuously becomes larger. In this sense, and up to the relatively small values of $r_{mis}$ for which the island ultimately still disappears ($W=0$ is still reached, see below), i.e.\ where the definition of \Eref{dtstablag} still yields finite values for $\Delta t_{stab}$, this time lag is expected to increase with increasing $r_{mis}$.

\begin{figure}
\includegraphics[scale=0.82]{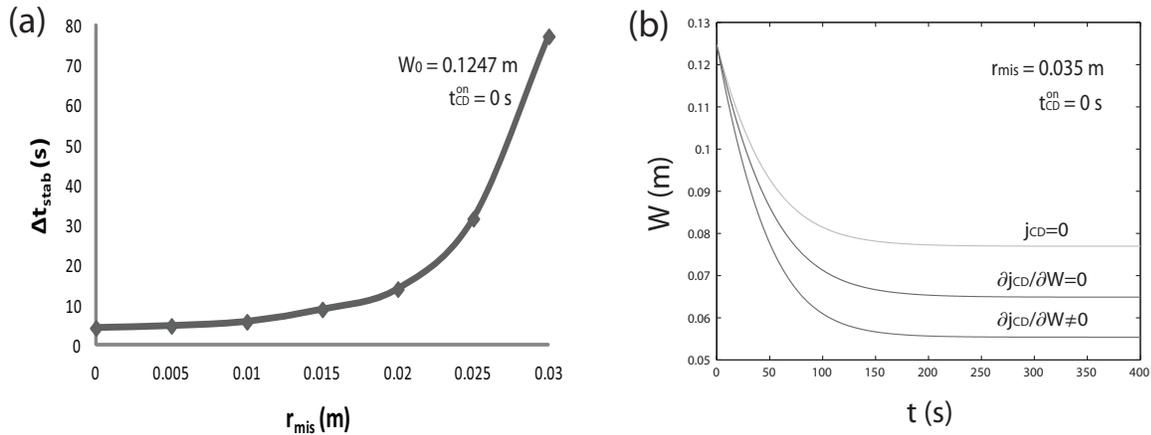}
\caption{Investigation of the dependence of the stabilization time lag on the radial misalignment of the EC beam with the island center: (a) $\Delta t_{stab}$ vs $r_{mis}$ for $W_0=0.1247$ m and $t^{on}_{CD}=0$ s, (b) $W$ vs $t$ for the same parameters and $r_{mis}=$ 0.035 m.}
\label{fig5}
\end{figure}

An additional computation of $\Delta t_{stab}$ for many different values of $r_{mis}$ within the range occurring in experiments ($< 3\,$cm) was performed, keeping the initial width constant at $W_0=0.1247\,$m and $t^{on}_{CD}=0\,$s. The results are plotted in \Fref{fig5}(a), and, according to these, the increase of $\Delta t_{stab}$ with increasing $r_{mis}$ is obvious. The scaling has been followed up to $r_{mis}= 3\,$cm, because for larger values of the misalignment the driven current is not optimized enough with the O-point position and the island  does ultimately not disappear anymore. This is verified in \Fref{fig5}(b), a plot similar to \Fref{fig3}, for the same parameters as in \Fref{fig5}(a) and with $r_{mis}= 3.5\,$cm, where the island stops shrinking at a width between 5.5 cm and 8 cm (depending on the simulation model). In this case, according to its definition, $\Delta t_{stab}$ becomes infinite, a trend that already appears in \Fref{fig5}(a).

The results presented so far on the deviation in the computed stabilization time between the two different models have been derived for a driven current based on the same injection power and geometry in all cases. A higher value of the current density achieved in terms of a larger value of $P_0$ would not yield a different picture for $\Delta t_{stab}$, since $\Delta^{'}_{CD}\propto j_{CD0}$ and the ratio of the ECCD peak values, as computed by the two models, would not alter because the island geometry encountered by the ray has not changed much. On the contrary, a higher value of $j_{CD0}$ achieved in terms of changing $\phi_l$ would result in a further increase of $\Delta t_{stab}$ since the ratio of the peak values increases due to geometric effects \cite{Isliker12}.

\section{Conclusion}
\label{conclusion}

In the context of the current drive requirements in modern fusion devices, including ITER and DEMO \cite{Poli12}, a self-consistent model for the dynamic evolution of the NTM growth in the presence of stabilizing ECCD has been performed, assuming linear wave-particle interaction and including the effect of the island geometry on the ECCD. The model includes the effect of the helical magnetic perturbation on the propagation, the flattening of the plasma electron pressure within the island region and the volumes of the perturbed flux surfaces into which the wave deposits it power. In general, the inclusion of the island geometry in the NTM dynamics has the problem of the distant time scales for the wave and the instability. The most efficient treatment would be to evolve the plasma process at certain time slices during the NTM evolution, and at each step to compute the ECCD profile with the ray tracing code and provide the results back to the plasma process. Here, we used an even simpler setting and avoided to use the wave code at each step, we numerically computed the current density for many values of the island width in advance, and then solved the MRE, getting the required values of the current density from the tabulated results of the wave computation.

Numerical computations were performed for the mode $3/2$, expected to be dominant in ITER, and parameters were chosen as relevant for a specific stabilization scenario, with using the upper EC launcher. The main results are the following:
\begin{itemize}
\item The mode stabilization occurs faster in terms of the self-consistent model as compared to the conventional
estimation, since the corresponding term $\Delta^{'}_{CD}$ in the MRE is always larger from the one in the axisymmetric case.
\item The time-lag $\Delta t_{stab}$ is an increasing function of the initial island width $W_0$, because the
geometric effects on $\Delta^{'}_{CD}$ are more important when the magnetic island is larger.
\item The scaling of $\Delta t_{stab}$ is almost independent of the EC turn-on time $t^{on}_{CD}$, since
$t^{on}_{CD}$ actually only causes a slight change of $W_0$.
\item $\Delta t_{stab}$ is an increasing function of $r_{mis}$, since for larger values of the misalignment
$\Delta^{'}_{CD}$ becomes larger with respect to the axisymmetric computation.
\end{itemize}

The fact that the stabilization process is computed to be faster when the island geometry is taken into account self-consistently in the ECCD suggests that the effect of the island geometry on the wave deposition is favorable for control, something that counteracts other known mechanisms that trim the ECCD efficiency and hamper the control effort, like e.g. quasilinear electron transport and wave beam broadening. The implementation of the island topology within the model allows for an accurate estimation of the ECCD effect on the NTM suppression. Revisiting the discussion in \cite{Isliker12}, it is verified that the enhanced ECCD peaking within the island may allow the use of less wave power than the one determined in axisymmetric geometry. This effect could serve in the direction of power economy, since less ECCD will be required for stabilization, provided that the coupling of power modulation with the frequency of island rotation is efficient.

A discussion of the limitations of our model is required. First, it has to be mentioned that the models for the wave propagation, absorption and current drive do not take into account diffraction or non-linear wave-particle interaction. Second, the coupling of the wave propagation with the island growth dynamics has been realized in terms of a linear model of the scaling of the driven current with the island width instead of a routinely use of the wave code at each step of the mode evolution. This increases the computational efficiency, and, within the linear plasma response context, it is sufficiently accurate. Third, the island dynamics were described in terms of a version of the modified Rutherford equation that includes a simplification in the description of the classical stability index $\Delta'$, which, in principle, should be  evaluated numerically by using the correct equilibrium current profile in a fully toroidal geometry. Finally, the stabilizing term $\Delta'_{CD}$ has been considered for the simple case where the island rotation is coped with by the ECRH system. The problem of asynchronism of the power modulation with the island rotation, as well as other relevant effects, are currently studied in the community (see e.g. \cite{Ayten12}).

Issues required to be studied further include the modeling of modulated and broad ECCD, which have been studied only partially in this article, and the inclusion of the effect of edge density fluctuations as a mechanism for undesired increase of the misalignment and/or the broadening of the EC beam. Also, a deeper study of the scaling laws of $\Delta t_{stab}$ with $W_0$ and $r_{mis}$ could provide valuable information in an effort to construct simple models for the effect of island topologies on the NTM dynamics. Other issues worthy of investigation are the modeling of the wave propagation with the plasma response computed in terms of the full particle dynamics in the non-axisymmetric fields, and the effect of the ECCD on the background magnetic equilibrium, which may have been underestimated in the computations done up to now.

\section{Acknowledgments}

The authors would like to thank Dr. A. Lazaros, Dr. E. Poli, Dr. E. Westerhof and B. Ayten for the very useful discussions. This work has been sponsored by the European Fusion Program (Association EURATOM-Hellenic Republic) and the Hellenic General Secretariat of Research and Technology. The sponsors do not bear any responsibility for the contents of this work.

\section*{References}

\bibliography{paper_230813}

\begin{thebibliography}{10}

\bibitem{Hegna97}
C.~C. Hegna and J.~D. Callen.
\newblock {\em Phys. Plasmas}, 4:2940, 1997.

\bibitem{Wilson08}
H.~R. Wilson.
\newblock {\em Trans. Fusion Science Tech.}, 53:152, 2008.

\bibitem{LaHaye06}
R.~J. LaHaye.
\newblock {\em Phys. Plasmas}, 13:055501, 2006.

\bibitem{Zohm07}
H.~Zohm, G.~Gantebein, F.~Leuterer, M.~Maraschek, E.~Poli, L.~Urso, and the
  ASDEX Upgrade~Team.
\newblock {\em Plasma Phys. Control. Fusion}, 49:B341, 2007.

\bibitem{Maraschek12}
M.~Maraschek.
\newblock {\em Nucl. Fusion}, 52:074007, 2012.

\bibitem{Urso09}
L.~Urso.
\newblock {\em Modelling and experiments on \textsc{NTM} stabilisation at
  \textsc{ASDEX} Upgrade}.
\newblock PhD thesis, Ludwig-Maximilians Universitaet Muenchen, 2009.

\bibitem{Rutherford73}
P.~H. Rutherford.
\newblock {\em Phys. Fluids}, 16:1903, 1973.

\bibitem{Ayten12}
B.~Ayten and E.~Westerhof.
\newblock {\em Phys. Plasmas}, 18:092506, 2012.

\bibitem{Rosa10}
P.~R. da~S.~Rosa and L.~F. Ziebell.
\newblock {\em Nucl. Fusion}, 50:115009, 2010.

\bibitem{Westerhof11}
E.~Westerhof.
\newblock {\em Nucl. Fusion}, 51:068001, 2011.

\bibitem{Tsironis09}
C.~Tsironis, A.~G. Peeters, H.~Isliker, D.~Strintzi, I.~Chatziantonaki, and
  L.~Vlahos.
\newblock {\em Phys. Plasmas}, 16:112510, 2009.

\bibitem{Lazaros07}
A.~Lazaros, M.~Maraschek, and H.~Zohm.
\newblock {\em Phys. Plasmas}, 14:042505, 2007.

\bibitem{Bernstein75}
I.~B. Bernstein.
\newblock {\em Phys. Fluids}, 18:320, 1975.

\bibitem{Friedland80}
L.~Friedland and I.~B. Bernstein.
\newblock {\em Phys. Rev. A}, 22:1680, 1980.

\bibitem{Mazzucato89}
E.~Mazzucato.
\newblock {\em Phys. Fluids B}, 1:1855, 1989.

\bibitem{Pereverzev98}
G.~V. Pereverzev.
\newblock {\em Phys. Plasmas}, 5:3529, 1998.

\bibitem{Westerhof89}
E.~Westerhof.
\newblock Implementation of \textsc{TORAY} at \textsc{JET}.
\newblock Technical Report 89-183, FOM Rijnhuisen, 1989.

\bibitem{Cohen87}
R.~H. Cohen.
\newblock {\em Phys. Fluids}, 30:2442, 1987.

\bibitem{Prater08}
R.~Prater, D.~Farina, Yu. Gribov, R.~W. Harvey, A.~K. Ram, Y.-R. Lin-Liu,
  E.~Poli, A.~P. Smirnov, F.~Volpe, E.~Westerhof, and A.~Zvonkov.
\newblock {\em Nucl. Fusion}, 48:035006, 2008.

\bibitem{Isliker12}
H.~Isliker, I.~Chatziantonaki, C.~Tsironis, and L.~Vlahos.
\newblock {\em Plasma Phys. Control. Fusion}, 54:095005, 2012.

\bibitem{Hamamatsu07}
K.~Hamamatsu, T.~Takizuka, N.~Hayashi, and T.~Ozeki.
\newblock {\em Plasma Phys. Control. Fusion}, 49:1955, 2007.

\bibitem{Lazzaro09}
E.~Lazzaro and S.~Nowak.
\newblock {\em Plasma Phys. Control. Fusion}, 51:035005, 2009.

\bibitem{Urso05}
L.~Urso, M.~Maraschek, H.~Zohm, and the ASDEX Upgrade~Team.
\newblock {\em J. Phys. Conf.}, 25:266, 2005.

\bibitem{Isayama07}
A.~Isayama, N.~Oyama, H.~Urano, T.~Suzuki, M.~Takechi, N.~Hayashi, K.~Nagasaki,
  Y.~Kamada, S.~Ide, T.~Ozeki, and the JT-60~team.
\newblock {\em Plasma Phys. Control. Fusion}, 47:773, 2007.

\bibitem{LaHaye02}
R.~J. LaHaye, S.~Guenter, D.~A. Humphreys, J.~Lohr, T.~C. Luce, M.~Maraschek,
  C.~C. Petty, R.~Prater, J.~T. Scoville, and E.~J. Strait.
\newblock {\em Phys. Plasmas}, 9:2051, 2002.

\bibitem{Kamendje05}
R.~Kamendje, S.~V. Kasilov, W.~Kernbichler, I.~V. Pavlenko, E.~Poli, and M.~F.
  Heyn.
\newblock {\em Phys. Plasmas}, 12:012502, 2005.

\bibitem{Sovinec04}
C.~R. Sovinec, A.~H. Glasser, T.~A. Gianakon, D.~C. Barnes, R.~A. Nebel, S.~E.
  Kruger, S.~J. Plimpton, A.~Tarditi, and M.~S. Chu.
\newblock {\em J. Comp. Phys.}, 195:355, 2004.

\bibitem{Jenkins10}
T.~G. Jenkins, S.~E. Kruger, C.~C. Hegna, D.~D. Schnack, and C.~R. Sovinec.
\newblock {\em Phys. Plasmas}, 17:012502, 2010.

\bibitem{Jenkins12}
T.~G. Jenkins and S.~E. Kruger.
\newblock {\em Phys. Plasmas}, 19:122508, 2012.

\bibitem{Stix92}
T.~H. Stix.
\newblock {\em Waves in Plasmas}.
\newblock Springer, 2nd edition, 1992.

\bibitem{Hazeltine92}
R.~D. Hazeltine and J.~D. Meiss.
\newblock {\em Plasma Confinement}.
\newblock Addison-Wesley, 1992.

\bibitem{Ren98}
C.~Ren, J.~D. Callen, T.~A. Gianakon, C.~C. Hegna, Z.~Chang, E.~D. Fredrickson,
  K.~M. McGuire, G.~Taylor, and M.~C. Zarnstorff.
\newblock {\em Phys. Plasmas}, 5:450, 1998.

\bibitem{Militello06}
F.~Militello, R.~J. Hastie, and F.~Porcelli.
\newblock {\em Phys. Plasmas}, 13:112512, 2006.

\bibitem{Arfken05}
G.~B. Arfken and H.~J. Weber.
\newblock {\em Mathematical Methods for Physicists}.
\newblock Elsevier, 6th edition, 2005.

\bibitem{Mikhailovskii03}
A.~B. Mikhailovskii.
\newblock {\em Contrib. Plasma Phys}, 43:125, 2003.

\bibitem{Poli12}
E.~Poli, E.~Fable, G.~Tardini, H.~Zohm, D.~Farina, L.~Figini, N.~B.
  Marushchenko, and L.~Porte.
\newblock {\em Plasma Phys. Control. Fusion}, 54:101766, 2012.

\end{thebibliography}

\end{document}